# Attitude Control of Spacecraft Swarms for Visual Mapping of Planetary Bodies

Ravi teja Nallapu
Space and Terrestrial Robotic Exploration (SpaceTREx) Laboratory
Dept. of Aerospace and Mechanical Engineering.
University of Arizona
Tucson, AZ 85721
rnallapu@email.arizona.edu

Jekanthan Thangavelautham,
Space and Terrestrial Robotic Exploration (SpaceTREx) Laboratory
Dept. of Aerospace and Mechanical Engineering.
University of Arizona
Tucson, AZ 85721
jekan @email.arizona.edu

*Abstract*— Planetary bodies such as asteroids, comets, and planetary moons are high-value science targets as they hold important information about the formation and evolution of our solar system. However, due to their low-gravity, variable sizes and shapes, dedicated orbiting spacecraft missions around these target bodies is difficult. Therefore, many planetary bodies are observed during flyby encounters, and consequently, the mapping coverage of the target body is limited. In this work, we propose the use of a spacecraft swarm to provide complete surface maps of a planetary body during a close encounter flyby. With the advancement of low-cost spacecraft technology, such a swarm can be realized by using multiple miniature spacecraft. The design of a swarm mission is a complex multi-disciplinary problem. To get started, we propose the Integrated Design Engineering & Automation of Swarms (IDEAS) software. In this work, we will introduce the development of the Automated Swarm Designer module of the software. The Automated Swarm Designer module will use evolutionary algorithms to optimize the design of swarms. The designed swarm will use 2 attitude control strategies to map the surface of a target body, namely: Nadir Pointing (NP), and Field of View Sweeping (FoV Sweeping). In the former strategy, the spacecraft are commanded to passively observe the sub-satellite point on the target body whenever the target is in the field of view of the individual spacecraft. This strategy is used when the observing instrument on board the spacecraft is large enough to capture the target from the desired encounter distance. In case the instrument is not large enough, then the spacecraft will have to maneuver their field of view to improve the coverage. The Field of View Sweeping strategy describes one such maneuver to improve coverage. In this strategy, the spacecraft in the swarm are commanded to sweep their fields of view about their principal axis normal to the swarm plane.

The current work addresses different aspects of the swarm design problem. First, the coverage problem is addressed, and the parameters: instrument size, and target flyby distance are determined. Following this, the modeling of the swarm is described, where the individual spacecraft are arranged to define a circular plane around the target body during the encounter. Then the encounter trajectories of the spacecraft swarm are modeled. Here a Newton-Raphson iterative scheme using the state transition matrix of the dynamics is proposed for the entire swarm, which finds the trajectories between the desired starting points and destination points of the spacecraft within a specified flyby duration. Following this, the two proposed attitude control strategies are then described as they occur on these flyby trajectories. Then the design of these swarms is presented using as 2 optimization problems. The proposed strategies are used to develop a numerical simulator which will serve as the Automated Swarm Designer module of IDEAS for visual mapping missions. Finally, the simulator developed is demonstrated by designing a swarm using genetic algorithms for a visual mapping mission of asteroid 433 Eros.

TABLE OF CONTENTS

1. INTRODUCTION

Exploration of small bodies and moons will provide insight into the origin of the solar system, the origins of Earth and the origin of life [1, 2]. The exploration of these bodies is also well supported by the latest planetary science decadal survey [3, 4]. These bodies are typically characterized by their small size, irregular shapes, and consequently irregular microgravity environments. Castillo-Rogez et al. [4] presents some of the popular scientific motivations for studying small bodies. In addition to these benefits, in-situ studies of near-Earth asteroids and planetary moons are also being pursued to provide resources for deep space travel [5]. While remote sensing observations from the ground provide useful information, they are limited by the resolution and atmospheric effects. For this reason, state-of-the-art missions to small bodies are targeting in-situ exploration. Several examples are shown in Figure 1 and include the OSIRIS-Rex, Hayabusa II, Rosetta and Deep Space 1.

Typically, orbiters and landers/rovers constitute the spacecraft architectures for exploring small bodies. However, the design of such spacecraft faces few key challenges: Firstly, the physical characteristics of these bodies are poorly understood, and therefore orbital reconnaissance of these targets is highly desired to derisk surface missions. Secondly, the dynamical motion of the spacecraft around the asteroid constrains the feasible orbits [6, 7]. Therefore, flyby observations are the ideal precursors for observing these small bodies. Typically, these flyby observations are carried by a single spacecraft equipped with the required instrument. However, the returns from a single observer spacecraft are limited by both the area of the target object accessible to the spacecraft instrument, and access duration. Additionally, a single spacecraft is highly susceptible to subsystem failures. Therefore, there is a strong motivation to search for better strategies to explore small bodies through flyby observations. Currently, miniature spacecraft of mass less than 50 kg are being developed as platforms to explore deep space [8, 9]. The total mission cost of these spacecraft is significantly less than their large competitors. For this reason, a swarm of multiple low-cost miniature spacecraft can be a viable option for exploring small bodies through flyby observations.

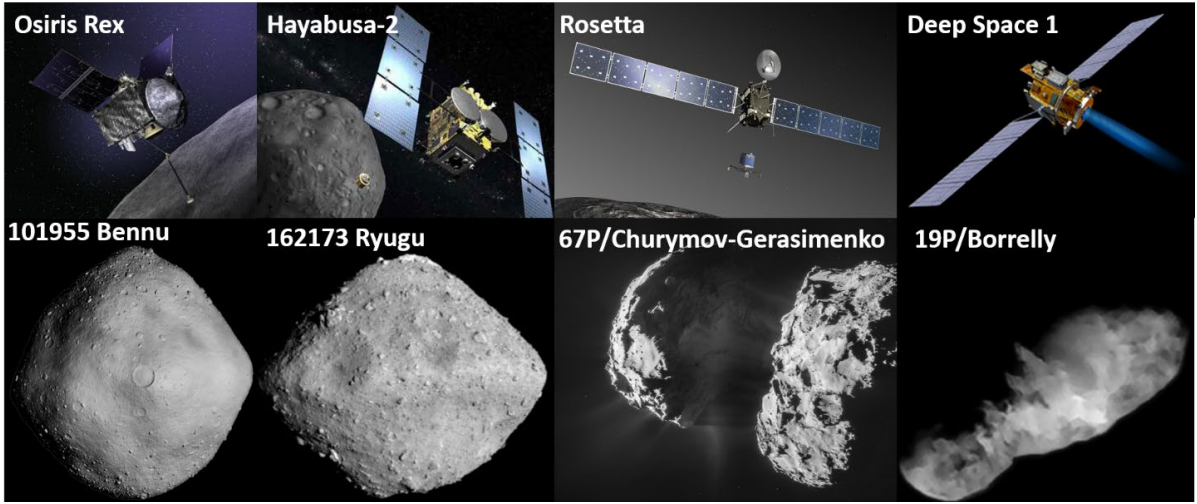

Figure 1: Well-known spacecraft missions for small body exploration. The spacecraft are shown on top, and their corresponding targets are shown below.

Typically, the design of a swarm-based mission involves selection of several parameters such as the number of spacecraft, choice of science payload, power system, communications, and propulsion. While each of these problems can be posed as decoupled design problems, an end-to-end design of such missions requires an integrated design approach. For this reason, we introduce the Integrated Design Engineering & Automation of Swarms (IDEAS), a software tool to design spacecraft swarm mission. The IDEAS software architecture is shown in Figure 2. As seen here, the software has 5 primary blocks: an input user interface; a knowledge base, a knowledge generator, a mission solver, and an output user interface. The mission designer defines the high-level mission requirements in the user input interface. The requirements placed can include parameters such as mission objectives, choice of the launch vehicle, the maximum and minimum number of spacecraft, and constraints on the spacecraft subsystems. These parameters are then sent to the mission solver which then solves each aspect of the design problem, namely trajectory, swarm design, and spacecraft design problem in a layered process. Alternately, the trajectory and spacecraft design maybe user inputs, in which case the solver attempts to optimize for the swarm configuration. The spacecraft trajectories and swarm operations are dynamical problems and involve a type of reference tracking using active or passive controls. These references tracked will be referred to as behaviors.

In this paper, we focus on developing the Automated Swarm Designer module of the IDEAS software. Using this module, we find optimal swarm configurations to perform flyby mapping of a target body. This work will focus on designing a spacecraft swarm with two attitude maneuvers to map the surface of a small body that includes (1) Nadir Pointing and (2) Field of View (FoV) sweeping. With Nadir Pointing, each spacecraft passively observes along their common center. With Field of View (FoV) Sweeping strategy, each spacecraft performs a sweep to maximize its effective field of view. The corresponding references for each attitude maneuver will be tracked by using a sliding mode control law. Using these observation methods, we develop a swarm system to perform maximum area coverage mapping of asteroid 433 Eros through flybys.

The paper is organized as follows: Section 2 presents related work. Section 3 presents the methodology used to design the swarm simulator to study the performance of swarms in small body missions. The trajectory finding and attitude tracking problems for individual spacecraft are also presented here. Section 4 presents a demonstration of the simulator. An example mission to map the surface of the asteroid 433 Eros with a desired ground resolution is presented. The results of the mapping simulations are presented here for a minimal sized swarm that can observe the complete surface of the target body. Finally, Section 5 presents conclusions and future work.

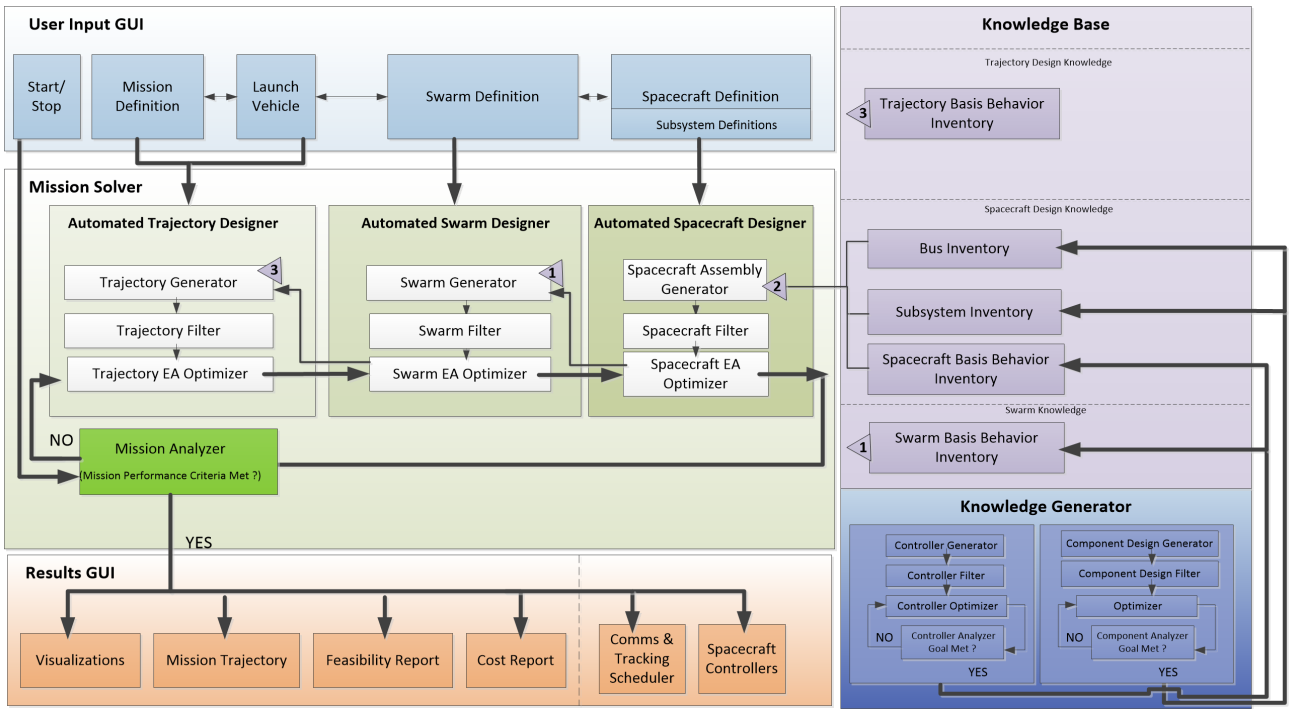

**Figure 2:** Software architecture of the proposed IDEAS software to provide an end-to-end design framework for spacecraft swarm missions.

## 2. RELATED WORK

Spacecraft swarms are being considered as a new space exploration platform. Although there is no well-defined bound on the number of spacecraft required to constitute a swarm, this work treats any collection of multiple spacecraft as a spacecraft swarm. This section will present the relevant research done in the field of multi-spacecraft technology. Multi-spacecraft missions are broadly classified into 2 types: formation flying and constellations [10]. Formation flying missions aim to couple the dynamics spacecraft so that they operate in some form of synchrony. Depending on the architecture, the formation flying missions are further classified into 2 types: centralized control [11], and decentralized control [12]. Centralized control architectures exhibit a central spacecraft, also known as the leader spacecraft, which computes the reference states and control laws of other participants in the swarm. While in the decentralized architectures, the constituent spacecraft make their own decisions. Constellations, on the other hand, require no coordination between their participants. Constellations have been successfully realized for Earth-based navigation [13], communications [14], with Earth-observation constellations in the works. Interplanetary constellations have also been proposed [15, 16]. Formation flying spacecraft missions have been studied for a wide range of mission types with objectives such as monitoring space weather [17], geodesy [18], gravity modelling [19, 20], deep space imaging [21], deep space exploration [22], on-orbit servicing [23], and distributed sensor networks [24, 25].

Spacecraft swarm architectures face several important challenges. The first is to maintain a required formation at times during a deep space mission to simplify communication and tracking with earth under the presence of environmental perturbations [10, 26]. Consequently, another challenge has been the inherent non-linearities developed in the dynamical modeling of the constituent spacecraft [27]. A third inherent challenge imposed on swarm technologies is the need to be cost-competitive with a large state-of-the-art monolithic spacecraft. The requirement to become a cost-effective platform for exploration may at times limit the capabilities of the constituent spacecraft [28].

The current state of the art research has focused on addressing these critical challenges. Modern research on spacecraft swarm guidance, navigation, and control (GNC) has focused on challenges such as the development of control laws for formation maintenance, robustness, cooperation, and swarm navigation. The formation maintenance problem is also known as the swarm keeping problem and has been well studied in the literature [26, 27, 29, 30, 31]. The research on swarm robustness focused on the algorithms to change the configurations of the spacecraft [29, 32, 33, 34]. Cooperation based research has focused on developing consensus-based algorithms for both maintenance and reconfiguration problems [35, 36]. Navigation research has focused on determining relative positions of the spacecraft in the swarm [37, 38].

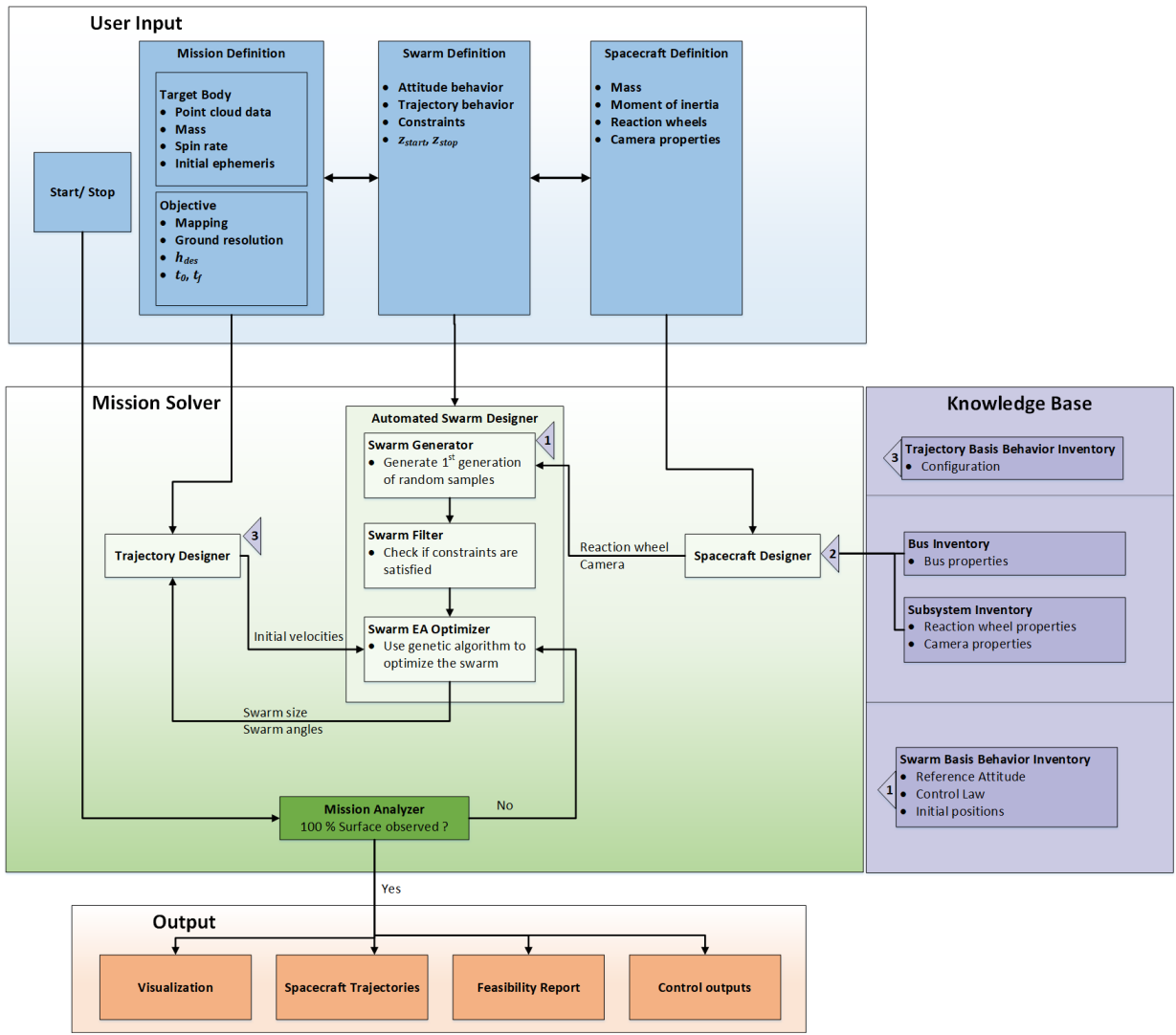

Figure 3: Detailed block diagram of Automated Swarm Designer, input parameters and connected modules.

Another line of research is the development of hardware platforms to realize and test swarm architectures. With the advent of subsystem technology for miniature spacecraft such as CubeSats [8, 16, 39], the feasibility of swarm architecture-based missions is rapidly increasing. Currently, platforms such as Chipsats from Cornell [40], SunCube FemtoSats from the University of Arizona [41, 42] and silicon wafer integrated Femtosats from JPL [43, 44] are being researched as hardware platforms for swarm-based space exploration.

While formation flying has many practical applications, it is not a requirement of a swarm. Applications such as global surface coverage/prospecting and persistent observations of target sites can be accomplished by an architecture that does not require the spacecraft to maintain relative positions. Therefore, such applications can be designed through swarm constellations. Constellation design research has focused on payload spatial and temporal coverage maximization [45, 46, 47]. In the current state of the art, constellations are designed using the grid point method [46] where the target region is specified by a grid of points on the target body surface. The performance of constellations of different shapes and structures are then tested either by varying them manually [48] or through a computer-based optimization scheme [49, 50]. However, constellation swarms have not been well studied for mapping of small bodies.

At the time of this work, spacecraft swarms have been considered as platforms to explore main belt asteroids utilizing distributed sensor networks [22, 51], mother-daughter swarm configurations [64] and as gravimetry platforms for asteroids through flybys [20, 52]. A major focus of asteroid exploration is the search for valuable resources such as water for spacecraft propulsion [65, 66]. The dynamics of spacecraft around irregular bodies such as asteroids and comets are being been well studied [6, 7, 53]. Relative equations of motion around irregular bodies for formation flying have been studied [54]. However, there is yet to be a unifying scheme for fast mapping of small bodies utilizing multiple spacecraft swarm flybys. This paper presents a new framework to design the trajectories of spacecraft swarms, where existing constellation design methods are combined with individual spacecraft motion to fly in a certain configuration.

## 3. Methodology

This section will describe the approach for designing different components of the swarm design simulator. This described simulator forms the Automated Swarm Designer module of the IDEAS software (see Figure 2). A more detailed block diagram identifying critical parameters is shown in Figure 3.

Firstly, we present the swarm configuration. In this approach, we present the swarms to be in a ring configuration. Other configurations include a helical configuration. Once the swarm has been configured, we then optimize the swarm in terms of the number of spacecraft to maximize area-coverage mapping using the NP and the FoV Sweeping maneuvers. Then we go further in depth to describe the 2 attitude behaviors: NP and FoV Sweeping along with the dynamics of the mapping.

*Swarm flyby*

For the application of mapping, we introduce a circular configuration of $N$ spacecraft around the asteroid arranged on the perimeter of a hypothetical ring as shown in Figure 4.

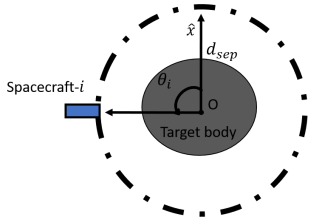

**Figure 4: Top view showing the ring configuration of the swarm around the target body.**

The radius of the ring is denoted as $d_{sep}$ which is expressed as:

$$d_{sep} = R_{av} + h_{des} \qquad (1)$$

Where $R_{av}$ is the average radius of the target body, and $h_{des}$ is the desired fly-by altitude of the spacecraft from the surface of the small body. The position of $i^{th}$ spacecraft on the ring is specified by its swarm angle $\theta_i$. We will assume that all the spacecraft move along the $z$-axis of the target body, which is typically its rotational pole as shown in Figure 5. The $x$ and $y$ axes can be chosen arbitrarily, but in the present work, the $x$-axis will be the direction pointing along the swarm angle $\theta_x = 0$ as shown in Figure 5.

*Trajectory design*—Since the spacecraft are assumed to move along $z$-axis in the rotating reference frame of the target body, only their $z$ coordinate with respect to the target body changes, while the $x$ and $y$ coordinates of the spacecraft stay fixed. Therefore, the position vector of the $i^{th}$ spacecraft with respect to the target body can be expressed as:

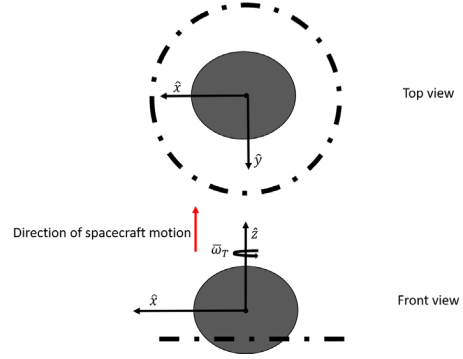

**Figure 5: Top and front views of the swarm configuration in the reference frame of the target body.**

$$\bar{r}_i(t) = \begin{bmatrix} d_{sep} \cos \theta_i \\ d_{sep} \sin \theta_i \\ z(t) \end{bmatrix} \qquad (2)$$

Let us assume that the spacecraft start initially at $z(t = t_0) = z_{start}$, and end up at $z(t = t_f) = z_{stop}$ as shown in Figure 6.

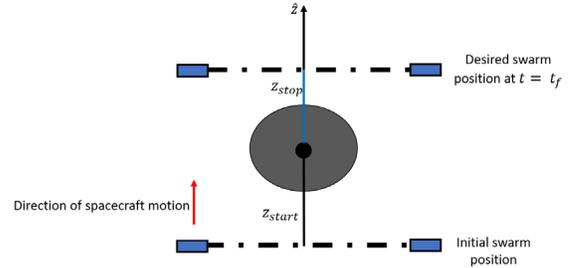

**Figure 6: Geometry of the flyby operations showing the starting and stopping locations of the swarm**

Now the trajectory design problem can be described as follows. To propagate a trajectory forward in time, the complete initial translational state information, i.e., $[\bar{r}_i(t_0) \; \bar{v}_i(t_0)]^T$ of all the spacecraft in the swarm are required. However, we now only have $\bar{r}_i(t_0)$ and $\bar{r}_i(t_f)$. Therefore, the problem now becomes: How to find the initial swarm velocities $\bar{v}_i(t_0)$ such the swarm reaches $\bar{r}_i(t_f)$ at $t = t_f$? –This is solved by using a Newton-Raphson iterative scheme on the state transition matrix which will now be presented.

*STM Targeting*—We begin by dividing the STM matrix into four $3 \times 3$ sub-matrices as follows:

$$\Phi(t, t_0) = \begin{bmatrix} \Phi_{rr} & \Phi_{vr} \\ \Phi_{rr} & \Phi_{vv} \end{bmatrix} \qquad (3)$$

In Equation 3, the submatrix $\Phi_{vr}$ corresponds to the sensitivity of the final position to the initial velocity. This will determine the correction term to a guessed initial velocity $\bar{v}_i(t_0)$. To use the method, we first begin with assuming an initial guess for the velocities $\bar{v}_i(0)$ as:

$$\bar{v}_{i0}(t_0) = \frac{\bar{r}_i(t_f) - \bar{r}_i(t_0)}{t_f - t_0} \quad (4)$$

We propagate the translational states forward in time until $t = t_f$ [53, 55] to obtain the states as $\bar{r}_{i0}(t_f)$, $\bar{v}_{i0}(t_f)$, and $\Phi_0(t_f, t_0)$. If the spacecraft experienced no net accelerations, Equation 4, would have resulted in a trajectory that reaches $\bar{r}_i(t_f)$ at $t = t_f$. However, perturbations such as Solar Radiation Pressure (SRP), and non-spherical gravity will induce an error in position:

$$\Delta \bar{r}_{i0} = \bar{r}_{i0}(t_f) - \bar{r}_i(t_f) \quad (5)$$

The details of these perturbations are discussed in [63]. From the propagated STM, we note the component velocity-position sensitivity component $\Phi_{vr,0}$. We now define an updated guess for the initial velocity as

$$\Delta \bar{v}_{i1} = \bar{v}_{i0}(t_f) - \Phi_{vr,0}^{-1}(t_f, t_0) \Delta \bar{r}_{i0} \quad (6)$$

This process is iterated till desired the trajectories converge to their corresponding $\bar{r}_i(t_f)$. For practical implementation purposes, we use 2 exit criteria on the iterative method described above: The first is to stop the iteration if the magnitude of the error during $k^{th}$ iteration falls below a tolerance parameter $\varepsilon$, ie, if $|\Delta \bar{r}_{ik}| \leq \varepsilon$. The second stopping condition occurs when the number of iterations exceed a set maximum number of iterations $N_{iter}$. This iteration is repeated for all the spacecraft in the 'ring' swarm. The trajectory design algorithm is summarized by the pseudocode shown in Figure 7.

```
Define: [Δr̄, v̄(t₀)] = propagate(r̄(t₀), v̄(t₀))
        Propagate to find r̄(t_f), v̄(t_f), and Φ(t_f, t₀);
        Note Φ_vr from Φ(t_f, t₀);
        Determine Δr̄ using Equation 5;
        Update v̄(t₀) using Equation 6;

For i = 1:N spacecrafts
    Determine v̄_{i0}(t₀) from Equation 4;
    [Δr̄_{i0}, v̄_{i0}(t₀)] = propagate(r̄_i(t₀), v̄_{i0}(t₀))
    For k = 0:N_iter iterations
        if |Δr̄_{ik}| ≤ ε
            break;
        [Δr̄_{ik+1}, v̄_{ik+1}(t₀)] = propagate(r̄_i(t₀), v̄_{ik}(t₀))
```

**Figure 7: Pseudocode to determine the initial conditions for the swarm**

*Attitude maneuvers*

Once the spacecraft trajectories are identified, the spacecraft in the swarm will execute attitude maneuvers to map the target body. This subsection will describe two such maneuvers used in the present work. We will proceed by describing the reference frames involved, and then describe the reference attitudes. In the current work, the attitude is propagated using the modified Rodriguez parameters (MRP) with shadow set switching to avoid controller unwinding and singularities [56].

*Swarm frame—* The swarm frame has its origin at the center of the ring swarm. As defined previously, the $x$-axis points along $\theta_x = 0$ and will be denoted by $\bar{r}_0$. The $z$-axis towards the closest vector to the average swarm velocity vector. The location and velocity of the center of the swarm are defined as follows:

$$R_{S,C} = \begin{bmatrix} 0 \\ 0 \\ \frac{\sum_{i=1}^{N} z_i}{N} \end{bmatrix} \quad (7)$$

$$\bar{v}_{S,C} = \begin{bmatrix} 0 \\ 0 \\ \frac{\sum_{i=1}^{N} v_{zi}}{N} \end{bmatrix} \quad (8)$$

The basis vectors of the swarm frame in the target body are described as follows:

$${}^S\hat{x} = \frac{\bar{r}_0 - \overline{OR}_{S,C}}{|\bar{r}_0 - \overline{OR}_{S,C}|} \quad (9)$$

$${}^S\hat{y} = \frac{\bar{v}_{S,C} \times {}^S\hat{x}}{|\bar{v}_{S,C} \times {}^S\hat{x}|} \quad (10)$$

and where:

$${}^S\hat{z} = {}^S\hat{x} \times {}^S\hat{y} \quad (11)$$

The rotation matrix that transforms the swarm frame into the target body frame is given by:

$$[TS] = [{}^S\hat{x} \quad {}^S\hat{y} \quad {}^S\hat{z}] \quad (12)$$

*Ring frames—* The ring frames describe the arrangement of the spacecraft inside the 'ring' swarm. The ring frames are defined such that the origin of the $i^{th}$ frame is located at the spacecraft center of mass, the $z$-axis points towards the center of the ring, the $x$-axis points in the direction of the $z$-axis of the swarm frame, and the $y$-axis is defined according to the right-hand thumb rule. Thus, the rotation matrix that transforms the swarm frame to the ring frame of the $i^{th}$ spacecraft is found as:

$$[R_i S] = \begin{bmatrix} 0 & 0 & -1 \\ 0 & 1 & 0 \\ 1 & 0 & 0 \end{bmatrix} \begin{bmatrix} 1 & 0 & 0 \\ 0 & \cos(\pi - \theta_i) & \sin(\pi - \theta_i) \\ 0 & -\sin(\pi - \theta_i) & \cos(\pi - \theta_i) \end{bmatrix} \quad (13)$$

The ring and swarm frames are illustrated in Figure 8. The left superscript $R$ corresponds to the vector resolved in the ring frame.

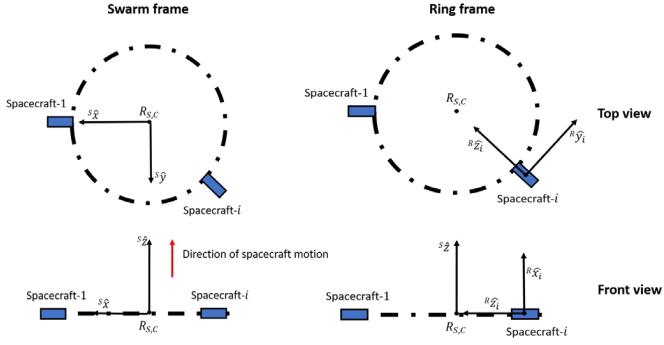

**Figure 8: Different frames associated with the swarm. The top and front views of the swarm frame are shown on the left, while those of the ring frame attached to $i^{th}$ spacecraft are shown on the right.**

Therefore, the rotation matrix that transforms the target body frame to the ring frame of the $i^{th}$ spacecraft is given by

$$[R_i T] = [R_i S][TS]^T. \quad (14)$$

In this work, the attitude of the $i^{th}$ spacecraft is defined as the orientation of its body frame with respect to its ring frame. With the reference frames established, the attitude maneuvers can now be explained.

*Swarm Optimization*

In order to make the design of the swarm automated, we will use an optimization routine in the loop in the current work. We will use 2 optimization problems corresponding to the 2 attitude maneuvers described in the current work. For the NP maneuver, we will determine a swarm with a minimum number of spacecraft such that the spacecraft observe 100 % surface of the target body. For the FoV Sweeping maneuver, as mentioned above, the FoV of the spacecraft will be half the FoV of the spacecraft in the NP simulation case. For this reason, we will use the same swarm but with half the FoV. However, we will try to determine the sweep periods. Therefore, the optimization problems used will be presented here.

Due to the discrete nature of these problems, genetic algorithms [57, 58] will be used to solve the optimization problems. The genetic algorithm optimization is probabilistic optimal solution search, which generates multiple designs or individuals, and through mechanisms analogous to evolution tries to locate an optimal solution. The search is carried through multiple generations, where designs that satisfy the objective function better have a better chance of selection.

*Nadir Pointing*— The design space of the swarm in the NP maneuver is expressed as a gene map as shown in Figure 16. The genes are used to describe a design to the genetic algorithms. As seen here, the design variables of the swarm in the NP maneuver is the swarm size and the corresponding swarm angles.

| Description | Swarm size | Spacecraft 1 Swarm angle | Spacecraft 2 swarm angle | ... | Spacecraft $N$ Swarm angle |
|---|---|---|---|---|---|
| Symbol | $N$ | $\theta_1$ | $\theta_2$ | | $\theta_N$ |
| Domain | Integers [1, $N_{max}$] | Real [0, $2\pi$] rad | | | |

**Figure 16: Gene map of the swarm used in determining the minimum number of spacecraft.**

The optimization problem in case of the NP problem can be expressed as

$$\min_{\substack{N, \ \theta_i \\ i=1:N}} N \quad (15)$$

Such that

$$\begin{aligned} |A_{Obs} - 100\ \%| - \epsilon_{Tol} &\leq 0 \\ 1 \leq N &\leq N_{max} \\ 0 \leq \theta_i &\leq 2\pi \end{aligned} \quad (16)$$

Where $A_{Obs}$ is the percentage of the surface area of the target body observed, $\epsilon_{Tol}$ is a tolerance parameter, and $N_{max}$ is the maximum number of spacecraft allowed. A value of $\epsilon_{Tol} = 0.1$ and $N_{max} = 20$ is used in the current work.

*FoV Sweeping*— The design space of the swarm in the FoV Sweeping maneuver is expressed as a gene map as shown in Figure 17. As seen here, the design variables of the swarm in the period of the sweep. The optimization problem in case of the FoV Sweeping problem can be expressed as

$$\max_{\substack{P_i \\ i=1:N}} A_{Obs} \quad (17)$$

Such that

$$P_{min} \leq P_i \leq P_{max} \quad (18)$$

| Description | Spacecraft 1 Sweep period | Spacecraft 2 Sweep period | ... | Spacecraft $N$ Sweep period |
|---|---|---|---|---|
| Symbol | $P_1$ | $P_2$ | | $P_N$ |
| Domain | Real [$P_{min}$, $P_{max}$] mins/sweep | | | |

**Figure 17: Gene map of the swarm used in designing the swarm using the FoV Sweeping maneuver.**

Where $P_{min}$ and $P_{max}$ are the minimum and maximum sweep periods the spacecraft can slew at. A lower $P_{min}$ implies faster sweeps are allowed, while larger $P_{max}$ would imply that slower sweeps are allowed. For the test spacecraft used in the current work a $P_{min} = 0.3$ mins/sweep was successfully tracked, and hence was the allowed minimum period. There was no upper bound on the maximum allowed period, i.e., $P_{max} = \infty$.

*Nadir Pointing*—The swarm will execute the NP maneuver when they have the full field of view ($\eta_N$) required to image the target body corresponding to a ground elevation angle $\varepsilon$. The field of view is determined using the sensor coverage relations described in [59]. We would like to recall here that the camera of the spacecraft is placed along its body $z$-axis. Therefore, in the NP maneuver, the spacecraft are required to align their body frame with respect to their ring frame as shown in Figure 12. The left superscript $B$ corresponds to the vector resolved in the body frame. In this case, the reference MRP, angular velocity and angular acceleration vectors are all zero.

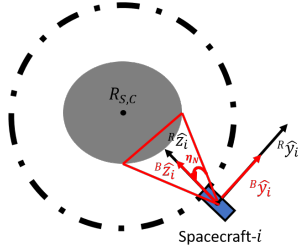

**Figure 12: Nadir Pointing maneuver showing that the spacecraft body frame aligns with its ring frame.**

*FoV Sweeping*—In the FoV Sweeping maneuver, the spacecraft perform a cross-track sweep of their field of view. It should be noted here that once a spacecraft with a half cone angle $\eta$ corresponding to an $\varepsilon$ moves away from its sub-satellite point $S$ by sweeping its FoV, the image obtained will not have a minimum ground elevation of $\varepsilon$ anymore. Therefore, in our current simulations, we assume that the spacecraft only has a half cone angle:

$$\eta_{FOV} = \frac{\eta_N}{2}. \qquad (19)$$

Which allows a maximum sweep angle of $\theta_{Max} = \frac{\eta_N}{2}$ clockwise, and counter clockwise and that would still allow a ground resolution of $\varepsilon$ as shown in Figure 13.

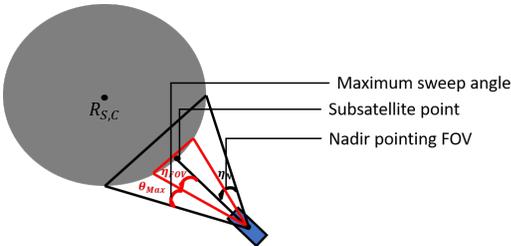

**Figure 13: Sweep geometry in FoV Sweep maneuver, showing the allowed FoV and the maximum sweep angle.**

Therefore, in the FoV Sweeping maneuver, the spacecraft in the swarm perform clockwise and counterclockwise sweep with a maximum angle $\theta_{Max}$ about their ring $x$-axis as shown in Figure 14. In order to facilitate an analytic formulation, the FoV Sweep is modelled as a sinusoidal oscillation with a sweeping period $P$. We first define the reference maneuver as a principal angle-principal axis pair [14], where the principal angle of $i^{th}$ spacecraft is given by

$$\theta_{i,R,FoV}(t) = \theta_{Max} \sin\left(\frac{2\pi}{P_i} t\right). \qquad (20)$$

And the principal axis is the $x$-axis, which is

$$\hat{e}_{R,FoV} = \begin{bmatrix} 1 \\ 0 \\ 0 \end{bmatrix}. \qquad (21)$$

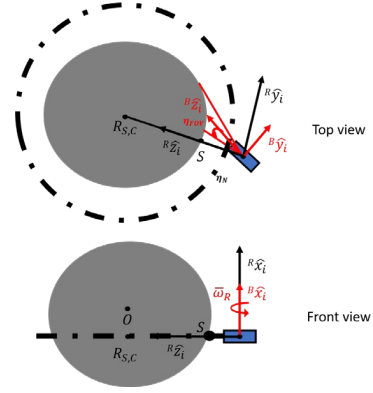

**Figure 14: Illustration of the FoV Sweeping maneuver.**

The reference angular velocities and angular accelerations can then be obtained by taking the time derivative of (20) in the $x$-axis:

$$\overline{\omega}_{R,FoV} = \theta_{Max}\left(\frac{2\pi}{P_i}\right) \cos\left(\frac{2\pi}{P_i} t\right) \begin{bmatrix} 1 \\ 0 \\ 0 \end{bmatrix}. \qquad (22)$$

and

$$\frac{d\overline{\omega}_{R,FoV}}{dt} = -\theta_{Max}\left(\frac{2\pi}{P_i}\right)^2 \sin\left(\frac{2\pi}{P_i} t\right) \begin{bmatrix} 1 \\ 0 \\ 0 \end{bmatrix}. \qquad (23)$$

Finally, the reference principal angle-principal vector pair is converted to the reference MRP using the conversion [56]:

$$\boldsymbol{\sigma}_{i,R,FoV} = \tan\left(\frac{\theta_{i,R,FoV}(t)}{4}\right) \hat{e}_{R,FoV} \qquad (24)$$

*Mapping operation*

This subsection describes the mapping operation done in the shape model of the target body to calculate the total mapped surface area. Two operations are done on the shape model to describe which vertices are observed, and which ones are unobserved. These are the culling and clipping operations which are as described below. Future versions of the simulator will include illumination-based filtering to map only those faces which are illuminated. However, in the current study, we assume all the faces are illuminated and hence can be mapped.

*Culling*—Typical asteroid shape models are available as a point cloud data [60] which contains a list of vertices, along with their vertex normal vectors. The culling operation lists the condition that the inner product of the line of sight vector of the spacecraft with respect to the target body $\hat{l}$, and the vertex normal of the $j^{th}$ vertex $\hat{n}_j$ is negative, where $\bar{l}_i \cdot \bar{n}_j \leq 0$ and $\bar{l}_i = -\bar{r}_i$. The culling operation on the shape model vertices is illustrated in Figure 15.

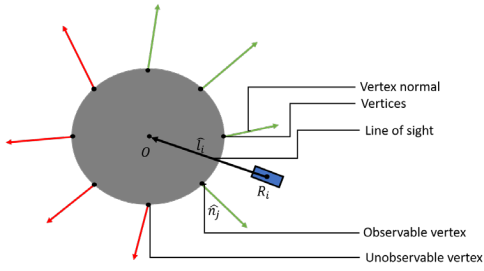

**Figure 15: Illustration of the culling operation.**

*Clipping*—Let $S_T$ denote the set of all vertices in the shape model of the target body, and $S_{Cull,i}(t)$ denote the set of all the vertices that were obtained after the culling operation using the line of sight of spacecraft-$i$ at time $t$. These vertices are then subjected to the camera transform [61]. We impose the clipping conditions such that a vertex $P_j = (a_j, b_j, c_j)$ in the image space will fall in the field of view of the spacecraft only if

$$\begin{bmatrix} |a_j| \\ |b_j| \\ |c_j| \end{bmatrix} \leq \begin{bmatrix} 1 \\ 1 \\ 1 \end{bmatrix} \quad (25)$$

All other points will be clipped-out from the shape model. Thus Equation 25 represents the clipping condition to determine the observed vertices of the shape model by a spacecraft.

If $S_{Clip,i}(t)$ denotes the set of all the vertices that fall in the FoV of spacecraft-$i$ at time $t$. The total observed surface area is determined by taking the set union over all spacecraft for the entire duration, i.e.,

$$S_{Obs} = \bigcup_{\substack{i=1:N \\ t=t_0:t_f}} S_{Clip,i}(t) \quad (26)$$

### 4. NUMERICAL SIMULATIONS

This section demonstrates the application of the swarm design simulator described above using an example mission. We will first design a swarm that uses the Nadir Pointing behavior using the optimization problem presented earlier. The trajectories of these swarms are propagated using the algorithm described in Figure 3.

*Mission objective*—As an example, we use a mission definition as follows: We are interested in obtaining a complete surface map of the asteroid 433 Eros with a ground resolution of 1 m and with a minimum ground elevation of 5 deg.

*Asteroid modeling*—A 1,700 polygon model of 433 Eros is used in this simulation. The parameters required for modeling the gravitational environment of Eros is listed in Table 1.

*Spacecraft modeling*—All the spacecraft in the swarm were assumed to be identical. The spacecraft were modeled using the parameters shown in Table 2. These parameters are used to form the Bus inventory module in Figure 2 and 3. The spacecraft mass was assumed to be distributed uniformly for calculating the moment of inertia tensor. The body axes were chosen to be the principal axes of the spacecraft, and the spacecraft was placed along the body $z$-axis, which is the axis of the least moment of inertia.

**Table 1. Model parameters for 433 Eros.**

| Parameter | Value |
|---|---|
| Mass | $6.689 \times 10^{15}$ kg |
| Rotation period | 5.270 hrs |
| $C_{20}$ | $-30.010$ km$^2$ |
| $C_{22}$ | 14.382 km$^2$ |
| Maximum radius | 17.670 km |
| Minimum radius | 3.222 km |
| Average radius | 10.447 km |
| Computed total surface area | 1103.342 km$^2$ |
| Heliocentric Semi-major axis | 1.457 AU |
| Eccentricity | 0.222 |
| Inclination | 10.828 deg |
| RAAN | 304.320 deg |
| Argument of periapsis | 178.820 deg |
| True anomaly | 0 deg |

**Table 2. Spacecraft parameters**

| Parameter | Value |
|---|---|
| Mass | $4 \; kg$ |
| Dimensions | $30 \times 10 \times 10 \; cm^3$ |
| Reflectivity | 0.6 |

*Instrument modeling*—The instrument model used in the simulations is presented in Table 3. It should be noted here that the near field distance $r_n$ isn't of much importance as long as it is small. The far field distance $r_f$ is chosen as the sum of the flyby altitude and twice the average radius of the target body.

**Table 3. Instrument parameters**

| Parameter | Value |
|---|---|
| Allowed aperture diameter | $8 \; cm$ |
| Imaging wavelength | 0.55 $\mu m$ |
| Minimum elevation angle | 5 $deg$ |
| Sensor half cone angle | 3.827 $deg$ |
| Near-field distance | 0.1 $mm$ |
| Far-field distance | 166.347 $km$ |
| Imaging frequency | 5 $images/sec$ |

*Flyby parameters*—The flyby parameters required to image Eros are determined from the data in Tables 1, 3, and from [59]. The z coordinates of the initial and final flyby trajectories are chosen as $\mp$ 30 km respectively. The duration of the trajectory is chosen as 10 min. The flyby parameters of the trajectory are summarized in Table 4.

**Table 4. Flyby parameters.**

| Parameter | Value |
|---|---|
| Flyby altitude | 145.454 km |
| $d_{sep}$ | 155.901 km |
| $d_{start}$ | −30 km |
| $d_{stop}$ | 30 km |
| $t_0$ | 0 min |
| $t_f$ | 10 min |
| Tolerance | 1 cm |
| Max iterations | 25 |

*Attitude control*—The spacecraft is assumed to be equipped with reaction wheels on all the 3-principal axis. As mentioned, the sliding mode controller in [62, 59] was used to track the reference attitudes for both the maneuvers described in Section 3. The parameters of the reaction wheels and the corresponding control gains used in the NP maneuver are presented in Table 5. The control gain parameters are defined using the same convention described in [62]. The camera properties and the attitude reaction wheel properties are used to define the Subsystem inventory module in Figure 18.

*Initial attitude*—The initial MRP of the spacecraft were chosen as a 3 × 1 vector with components that are uniformly distributed random numbers in [−1, 1], which were then switched to their corresponding shadow sets [56], in case their magnitude exceeded unity. The initial conditions for the spacecraft angular velocity and reaction wheel angular velocities were chosen as random 3 × 1 vectors with components in [−1, 1] RPM and [−10, 10] RPM respectively. The integration time step was chosen as half of the imaging frequency mentioned in Table 3 (i.e., the time step used is 0.1 s).

**Table 5. Attitude actuator and control parameters during the Nadir Pointing maneuver**

| Parameter | Value |
|---|---|
| Reaction wheel mass | 240 g |
| Reaction wheel radius | 2.9 cm |
| Maximum torque | 0.007 Nm |
| Maximum angular momentum | 0.05 Nms |
| $K_P$ | 0.4 rad/s |
| $K_I$ | 0.01 rad/s$^2$ |
| $\eta_g$ | 0.5 rad/s$^2$ |
| $\phi_g$ | 100 rad/s |

*Swarm sizing*—The optimization problem in Equations 15 and 16 is solved using a mixed integer genetic algorithm [58]. The simulation was run for 115 generations at which point the solution converged to a swarm size of 8 to attain 100% area coverage of the target body (Figure 19). The final generation of the evolutionary search run contained 100 solutions shown in Figure 20.

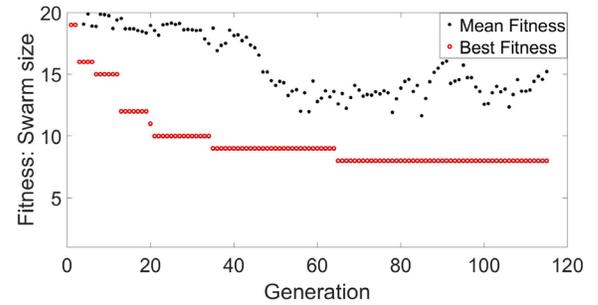

**Figure 19: Population mean fitness and population best fitness during evolutionary search run using Nadir Pointing for 110 generations.**

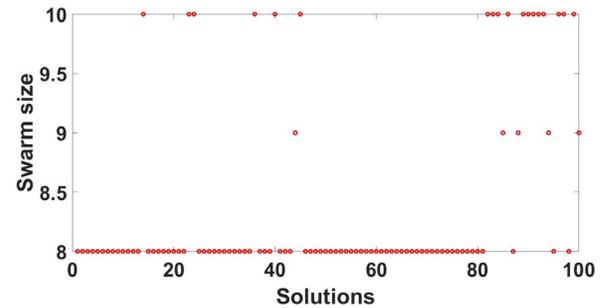

**Figure 20: Results of the swarm size optimization problem, showing the variation of the swarm size of the among the solutions in the final generation.**

As seen here, a swarm consisting of $N = 8$ spacecraft was the fewest number of spacecraft that achieves the required 100 % coverage of Eros. The final solution selected is shown in Figure 21 in the gene map format.

| Swarm size | Spacecraft 1 Swarm angle | Spacecraft 2 Swarm angle | Spacecraft 3 Swarm angle | Spacecraft 4 Swarm angle | Spacecraft 5 Swarm angle | Spacecraft 6 Swarm angle | Spacecraft 7 Swarm angle | Spacecraft 8 Swarm angle |
|---|---|---|---|---|---|---|---|---|
| 8 | 0.2 | 1.7 | 2.4 | 2.9 | 3.3 | 3.5 | 5.6 | 6.0 |

**Figure 21: Selected solution of the swarm that uses the NP maneuver.**

*Trajectories*—The trajectories of the swarm propagated using the algorithm are shown in Figure 22, and the associated error $|\Delta \bar{r}_{ik}|$ is shown in Figure 23. As mentioned in Table 4, a maximum of $N_{iter} = 25$ iterations, with a tolerance of 1 cm. The time span was chosen to be between [0,10] mins.

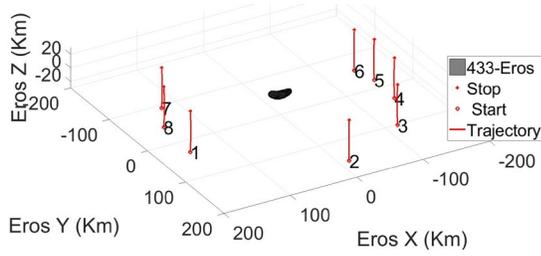

Figure 22: Generated trajectories of the 8-spacecraft ring swarm.

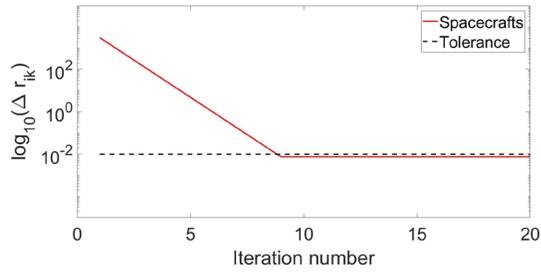

Figure 23: Variation of error magnitude $|\Delta \bar{r}_{ik}|$ with iteration count for all spacecraft.

*Nadir Pointing*—The flyby simulation of the selected swarm near 433-Eros is shown in Figure 24. As expected all the 8 spacecraft face their FoV pyramids towards their ring center.

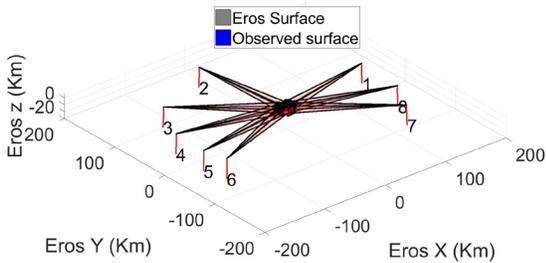

Figure 24: Nadir Pointing maneuver showing the FoV pyramids pointed towards the ring center.

The surface of Eros after the flyby operation, i.e., at $t = t_f$ is shown in Figure 25. The surface shaded in blue is formed by all the vertices that satisfy Equation 26. As expected the swarm is able to observe the complete asteroid surface if the entire surface is illuminated.

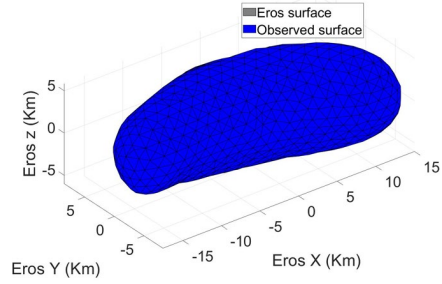

Figure 25: Surface of Eros after the Nadir Pointing maneuver of the swarm, showing 100 % surface coverage.

The attitude errors of the spacecraft in the ring swarm are shown in Figure 26. The MRP error shown here is computed using the MRP difference described in [58]. The MRP tracking errors are shown to asymptotically approach 0. The angular velocity tracking errors of the spacecraft in the ring swarm are shown in Figure 27. Spacecraft track the reference angular velocity as expected. The reaction wheel angular velocities of the spacecraft are shown in Figure 28 and the reaction wheel control torques of all the spacecraft are shown in Figure 29. The reaction wheel spin rates of all the spacecraft are expected to be within the specified bounds. The control torque outputs from the reaction wheels of the spacecraft are well within the specified bounds.

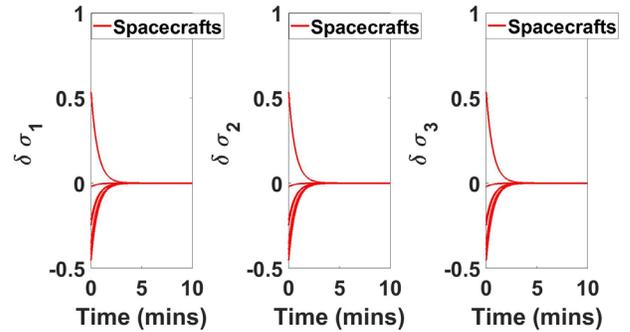

Figure 26: MRP tracking errors of the spacecraft in the ring swarm during the Nadir Pointing maneuver.

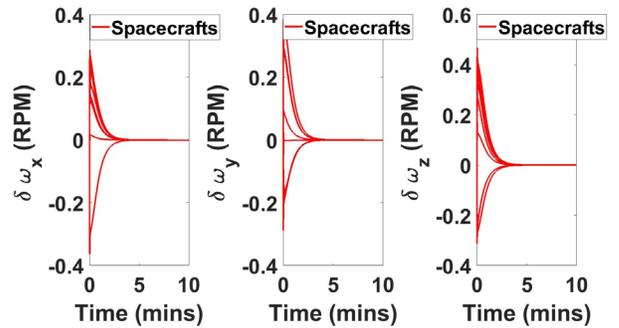

Figure 27: Spacecraft angular velocity tracking errors in the ring swarm during the Nadir Pointing maneuver.

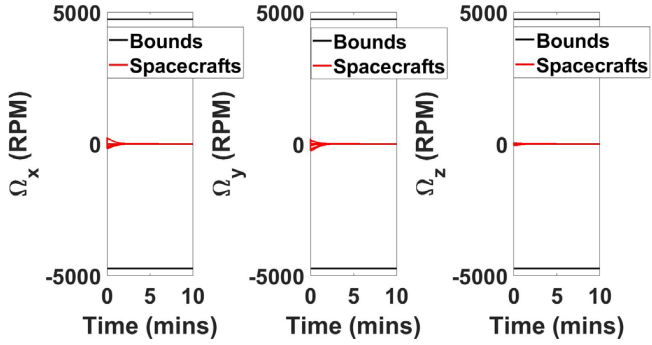

Figure 28: Reaction wheel spin rates during the Nadir Pointing maneuver.

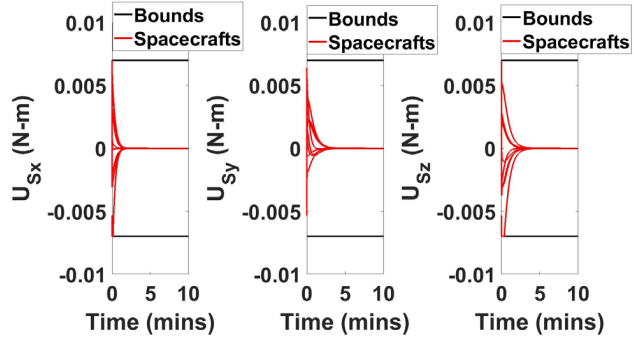

Figure 29: Control torques exerted by the reaction wheels during the Nadir Pointing maneuver.

*FoV Sweeping*—As described earlier, in the FoV Sweeping simulations, the spacecraft have a half cone angle of $\eta_{FOV} = \frac{\eta_N}{2}$, and must sweep a maximum angle of $\frac{\eta_N}{2}$. In this case, an aggressive controller was used to track the reference attitude. The parameters for simulating the FoV Sweeping maneuver are listed in Table 6.

**Table 6. Instrument and control parameters used to simulate the FoV Sweeping maneuver.**

| Parameter | Value |
| --- | --- |
| Minimum elevation angle | 5 deg |
| Sensor half cone angle | 1.9138 deg |
| Maximum sweep angle | 1.9138 deg |
| Sweep period | 30 s |
| $K_P$ | [4 0.4 0.4] rad/s |
| $K_I$ | [1 0.01 0.01] rad/s$^2$ |

The sweep periods are determined by solving the optimization problem presented in Equations 17 and 18. The optimization ran for 51 generations which evaluated 10,400 different designs (Figure 30). As seen, the solution converged to 100% area coverage. The final generation contained 200 solutions. The area observed by the solutions in the final generation is presented in Figure 31. As seen here, most solutions are able to observe 100 % surface of the asteroid. The selected solution is presented in the gene format in Figure 32. The flyby simulation of the swarm when executing the FoV Sweep maneuver is shown in Figure 33.

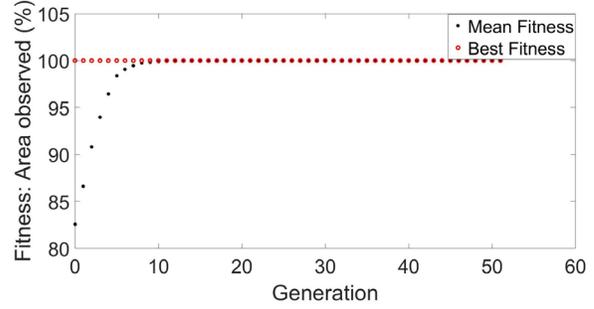

Figure 30: Population mean fitness and population best fitness during evolutionary search run using FoV sweep optimization.

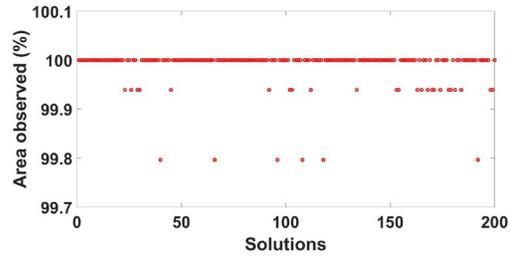

Figure 31: Results of the sweep rate optimization problem, showing how the variation of the coverage area of the by the solutions in the final generation.

| Spacecraft 1 Sweep period | Spacecraft 2 Sweep period | Spacecraft 3 Sweep period | Spacecraft 4 Sweep period | Spacecraft 5 Sweep period | Spacecraft 6 Sweep period | Spacecraft 7 Sweep period | Spacecraft 8 Sweep period |
| --- | --- | --- | --- | --- | --- | --- | --- |
| 0.6 | 0.3 | 0.3 | 7.4 | 0.5 | 0.7 | 1.2 | 2.3 |

Figure 32: Selected optimal solution of the swarm that uses the FoV Sweeping maneuver. The solution is expressed in the gene map format.

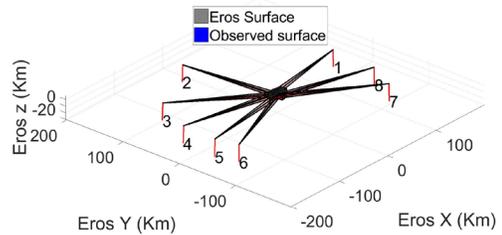

Figure 33: The FoV Sweeping maneuver showing the FoV pyramids perform a cross track sweep.

The MRP tracking errors of the spacecraft during the FoV Sweeping maneuver are shown in Figure 34 and the angular velocity tracking errors are shown in Figure 35 where they are shown to track their reference attitudes. The reaction wheel spin rates of the spacecraft are shown in Figure 36 and the control torques from the reaction wheels are shown in Figure 37. As observed here, the reaction wheel spin rates and their control torques are observed to be well within the specified bounds.

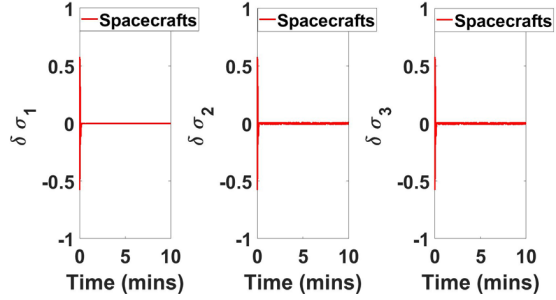

Figure 34: MRP tracking errors of the spacecraft in the ring swarm during the FoV Sweeping maneuver.

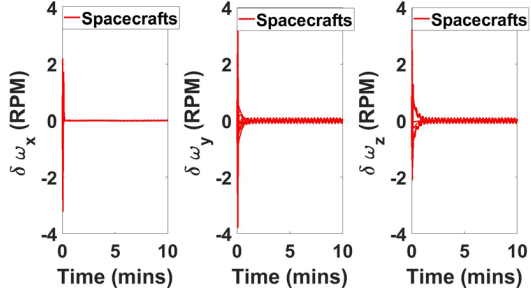

Figure 35: Spacecraft angular velocity tracking errors in the ring swarm during the FoV Sweeping maneuver.

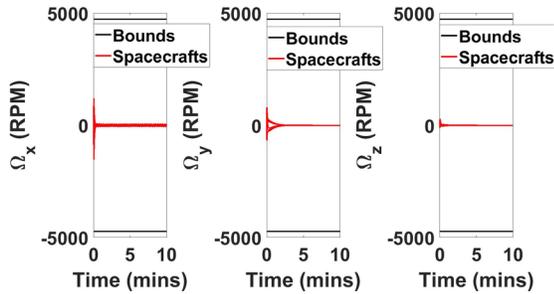

Figure 36: Reaction wheel spin rates during the FoV Sweeping maneuver.

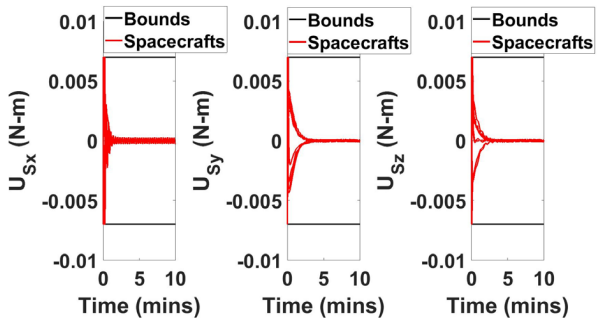

Figure 37: Reaction wheel control torques of the spacecraft in the 'ring' swarm during the FoV Sweeping maneuver.

5. DISCUSSION

The results of the mapping simulations presented in this section emphasize two key points. The first is the advantage of the swarms. Evidently, a 100 % coverage of the target body during a flyby is not possible by a single spacecraft due to its limited access area. Utilizing current state-of-art mapping techniques, a single spacecraft requires that it enters orbit around the target body or use additional fuel to alter their trajectories to do multiple flybys which are all very expensive. As mentioned earlier, orbits around these target bodies are challenging due to the dynamical environment. Therefore, this would complicate the spacecraft design. The use of additional fuel to map the target bodies also increases the spacecraft mass, and thus cannot be used as a generalized approach to map the 700,000+ asteroids discovered in the Solar system. Using a swarm, on the other hand, we can map the target body in one pass as shown in this work.

The second point that the results highlight is that with a swarm architecture, exploration tasks can be automated as behaviors. The NP and FoV Sweeping maneuvers are two autonomous examples. This now opens pathways to more advanced maneuvers for efficient mapping and small body exploration.

6. CONCLUSIONS

This paper presents a new, automated method of exploring small bodies utilizing the IDEAS software framework. The work uses the Automated Swarm Design module that utilizes Evolutionary Algorithms to design and configure a spacecraft swarm that uses 2 attitude control behaviors: Nadir Pointing (NP) and the Field of View (FoV) Sweeping to completely map the surface of the target body at a specified ground resolution by performing a coordinated flyby. The software tunes the relative positions of the spacecraft in the swarm to maximize area-coverage mapping of odd-shaped small-bodies which is a non-trivial task. A coordinated flyby is expected to be cost-effective to map a small-body than getting into orbit. A numerical simulation of the complete mapping is demonstrated for asteroid 433-Eros with a desired ground resolution of 1 m, and a minimum ground elevation angle of 5 deg. The corresponding instrument and flyby parameters are found from the relations described in the current work, and it is shown that a swarm of 8 spacecraft is able to obtain 100 % coverage of Eros in one flyby when using the NP maneuver. This can have applications in thermal mapping. Future studies using the simulator will also factor in the illumination from the Sun when designing visual mapping missions.

state estimation of spacecraft near small Solar System bodies. *Advances in Space Research*, *57*(8), pp.1747-1761.
[56] Junkins, J.L. and Schaub, H., 2014. *Analytical mechanics of space systems*. American Institute of Aeronautics and Astronautics.
[57] Coello, C.A.C., Lamont, G.B. and Van Veldhuizen, D.A., 2007. *Evolutionary algorithms for solving multi-objective problems* (Vol. 5). New York: Springer.
[58] Conn, A.R., Gould, N.I. and Toint, P., 1991. A globally convergent augmented Lagrangian algorithm for optimization with general constraints and simple bounds. *SIAM Journal on Numerical Analysis*, *28*(2), pp.545-572.
[59] Wertz, J.R., Everett, D.F. and Puschell, J.J., 2011. *Space mission engineering: the new SMAD*. Microcosm Press.
[60] Space Frieger Shape model catalogue, "https://space.frieger.com/asteroids/ 2018".
[61] Joy, Ken. "Camera Transform." *ECS 178 - Geometric Modeling Transformations*, UC Davis, 2012.
[62] Kowalchuk, S. and Hall, C., 2008, August. Spacecraft attitude sliding mode controller using reaction wheels. In *AIAA/AAS Astrodynamics Specialist Conference and Exhibit* (p. 6260).
[63] Nallapu, R., "User Guide for the IDEAS Spacecraft Swarm Design Software," *Space and Terrestrial Robotic Exploration Laboratory Technical Report*, 2019.
[64] Vance, L., Asphaug, E., Thangavelautham, J., "Evaluation of Mother-Daughter Architectures for Asteroid Belt Exploration," *AIAA Science and Technology Forum*, pp. 1-8, 2019.
[65] Pothamsetti, R., Thangavelautham, J.,"Photovoltaic Electrolysis Propulsion System for Interplanetary CubeSats," *Proceedings of the IEEE Aerospace Conference*, 2016.
[66] Rabade, S., Barba, N., Garvie, L., Thangavelautham, J., "The Case for Solar Thermal Steam Propulsion System for Interplanetary Travel: Enabling Simplified ISRU Utilizing NEOs and Small Bodies," Proceedings of the 67th International Astronautical Congress, 2016.

BIOGRAPHY

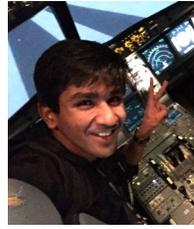

***Ravi Teja Nallapu*** received a B.Tech. in Mechatronics Engineering from JNTU, Hyderabad, India in 2010. He then received an M.S in Aerospace Engineering from the University of Houston, TX in 2012. After this, he worked with U.S Airways as a Flight Simulator Engineer for 2 years. He is presently pursuing his Ph.D. in Aerospace Engineering from the University of Arizona, Az. He specializes in control theory and robotics. His research interests include GNC of spacecraft swarms, space systems engineering, orbital mechanics, and exploration robotics.

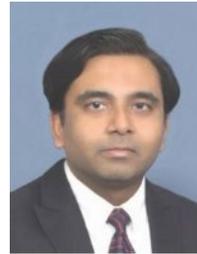

***Jekan Thangavelautham*** *is an Assistant Professor and has a background in aerospace engineering from the University of Toronto. He worked on Canadarm, Canadarm 2 and the DARPA Orbital Express missions at MDA Space Missions. Jekan obtained his Ph.D. in space robotics at the University of Toronto Institute for Aerospace Studies (UTIAS) and did his postdoctoral training at MIT's Field and Space Robotics Laboratory (FSRL). Jekan Thanga heads the Space and Terrestrial Robotic Exploration (SpaceTREx) Laboratory at the University of Arizona. He is the Engineering Principal Investigator on the AOSAT I CubeSat Centrifuge mission and is a Co-Investigator on SWIMSat, an Airforce CubeSat mission to monitor space threats.*

16